\input harvmac
\let\includefigures=\iftrue
\let\useblackboard==\iftrue
\newfam\black

\includefigures
\message{If you do not have epsf.tex (to include figures),}
\message{change the option at the top of the tex file.}
\def\figin{\epsfcheck\figin}\def\figins{\epsfcheck\figins}
\def\epsfcheck{\ifx\epsfbox\UnDeFiNeD
\message{(NO epsf.tex, FIGURES WILL BE IGNORED)}
\gdef\figin##1{\vskip2in}\gdef\figins##1{\hskip.5in}
\else\message{(FIGURES WILL BE INCLUDED)}%
\gdef\figin##1{##1}\gdef\figins##1{##1}\fi}
\def\DefWarn#1{}
\def\figinsert{\goodbreak\midinsert}
\def\ifig#1#2#3{\DefWarn#1\xdef#1{fig.~\the\figno}
\writedef{#1\leftbracket fig.\noexpand~\the\figno}%
\figinsert\figin{\centerline{#3}}\medskip\centerline{\vbox{
\baselineskip12pt\advance\hsize by -1truein
\noindent\footnotefont{\bf Fig.~\the\figno:} #2}}
\endinsert\global\advance\figno by1}
\else
\def\ifig#1#2#3{\xdef#1{fig.~\the\figno}
\writedef{#1\leftbracket fig.\noexpand~\the\figno}%
\global\advance\figno by1} \fi
\def\id{{1 \kern-.28em {\rm l}}}

\def\K3{{\bf K3}}
\def\journal#1&#2(#3){\unskip, \sl #1\ \bf #2 \rm(19#3) }
\def\andjournal#1&#2(#3){\sl #1~\bf #2 \rm (19#3) }

\def\bar{\overline}

\def\ie{{\it i.e.}}
\def\eg{{\it e.g.}}

\def\tilde{\widetilde}

\def\frac#1#2{{#1\over#2}}

\def\inbar{\,\vrule height1.5ex width.4pt depth0pt}
\def\IC{\relax\hbox{$\inbar\kern-.3em{\rm C}$}}
\def\IR{\relax{\rm I\kern-.18em R}}
\def\IZ{\relax{\rm I\kern-.18em Z}}

%
%

%
\catcode`\@=11
\def\slash#1{\mathord{\mathpalette\c@ncel{#1}}}
\overfullrule=0pt

\def\FF{{\cal F}}

\def\HH{{\cal H}}

\def\LL{{\cal L}}
\def\MM{{\cal M}}
\def\NN{{\cal N}}

\def\ZZ{{\cal Z}}

\def\underrel#1\over#2{\mathrel{\mathop{\kern\z@#1}\limits_{#2}}}

\catcode`\@=12


%

\def\exp{{\rm exp}}


\def\ie{{\it i.e.}}
\def\eg{{\it e.g.}}


\lref\HashimotoWCT{
  A.~Hashimoto and D.~Kutasov,
  ``$T \bar{T},J \bar T$, $T \bar{J}$ Partition Sums From String Theory,''
[arXiv:1907.07221 [hep-th]].
}
\lref\GiveonNIE{
  A.~Giveon, N.~Itzhaki and D.~Kutasov,
  ``$ T \bar T $ and LST,''
JHEP {\bf 1707}, 122 (2017).
[arXiv:1701.05576 [hep-th]].
}
\lref\DijkgraafXW{
  R.~Dijkgraaf, G.~W.~Moore, E.~P.~Verlinde and H.~L.~Verlinde,
  ``Elliptic genera of symmetric products and second quantized strings,''
Commun.\ Math.\ Phys.\  {\bf 185}, 197 (1997).
[hep-th/9608096].
}
\lref\KlemmDF{
  A.~Klemm and M.~G.~Schmidt,
  ``Orbifolds by Cyclic Permutations of Tensor Product Conformal Field Theories,''
Phys.\ Lett.\ B {\bf 245}, 53 (1990).
}
\lref\AtickSI{
  J.~J.~Atick and E.~Witten,
 ``The Hagedorn Transition and the Number of Degrees of Freedom of String Theory,''
Nucl.\ Phys.\ B {\bf 310}, 291 (1988).
}

\lref\GiveonZM{
  A.~Giveon, D.~Kutasov and O.~Pelc,
  ``Holography for noncritical superstrings,''
JHEP {\bf 9910}, 035 (1999).
[hep-th/9907178].
}
\lref\CardySDV{
  J.~Cardy,
  ``The $ T\overline{T} $ deformation of quantum field theory as random geometry,''
JHEP {\bf 1810}, 186 (2018).
[arXiv:1801.06895 [hep-th]].
}
\lref\DubovskyBMO{
  S.~Dubovsky, V.~Gorbenko and G.~Hernández-Chifflet,
  ``$ T\overline{T} $ partition function from topological gravity,''
JHEP {\bf 1809}, 158 (2018).
[arXiv:1805.07386 [hep-th]].
}
\lref\AharonyBAD{
  O.~Aharony, S.~Datta, A.~Giveon, Y.~Jiang and D.~Kutasov,
  ``Modular invariance and uniqueness of $T\bar{T}$ deformed CFT,''
JHEP {\bf 1901}, 086 (2019).
[arXiv:1808.02492 [hep-th]].
}
\lref\AharonyICS{
  O.~Aharony, S.~Datta, A.~Giveon, Y.~Jiang and D.~Kutasov,
  ``Modular covariance and uniqueness of $J\bar{T}$ deformed CFTs,''
JHEP {\bf 1901}, 085 (2019).
[arXiv:1808.08978 [hep-th]].
}
\lref\ChakrabortyMDF{
  S.~Chakraborty, A.~Giveon and D.~Kutasov,
  ``$T\bar{T}$, $J\bar{T}$, $T\bar{J}$ and String Theory,''
J.\ Phys.\ A {\bf 52}, no. 38, 384003 (2019).
[arXiv:1905.00051 [hep-th]].
}
\lref\LeFlochRUT{
  B.~Le Floch and M.~Mezei,
  ``Solving a family of $T\bar{T}$-like theories,''
[arXiv:1903.07606 [hep-th]].
}

\lref\ChakrabortyKPR{
  S.~Chakraborty, A.~Giveon, N.~Itzhaki and D.~Kutasov,
  ``Entanglement beyond AdS,''
Nucl.\ Phys.\ B {\bf 935}, 290 (2018).
[arXiv:1805.06286 [hep-th]].
}

\lref\MaldacenaUZ{
  J.~M.~Maldacena, J.~Michelson and A.~Strominger,
  ``Anti-de Sitter fragmentation,''
JHEP {\bf 9902}, 011 (1999).
[hep-th/9812073].
}

\lref\KutasovXU{
  D.~Kutasov and N.~Seiberg,
  ``More comments on string theory on AdS(3),''
JHEP {\bf 9904}, 008 (1999).
[hep-th/9903219].
}

\lref\ChakrabortyVJA{
  S.~Chakraborty, A.~Giveon and D.~Kutasov,
  ``$ J\overline{T} $ deformed CFT$_{2}$ and string theory,''
JHEP {\bf 1810}, 057 (2018).
[arXiv:1806.09667 [hep-th]].
}

\lref\DijkgraafVV{
  R.~Dijkgraaf, E.~P.~Verlinde and H.~L.~Verlinde,
  ``Matrix string theory,''
Nucl.\ Phys.\ B {\bf 500}, 43 (1997).
[hep-th/9703030].
}

\lref\GiveonMYJ{
  A.~Giveon, N.~Itzhaki and D.~Kutasov,
  ``A solvable irrelevant deformation of AdS$_{3}$/CFT$_{2}$,''
JHEP {\bf 1712}, 155 (2017).
[arXiv:1707.05800 [hep-th]].
}
\lref\AharonyUB{
  O.~Aharony, M.~Berkooz, D.~Kutasov and N.~Seiberg,
  ``Linear dilatons, NS five-branes and holography,''
JHEP {\bf 9810}, 004 (1998).
[hep-th/9808149].
}

\lref\SmirnovLQW{
  F.~A.~Smirnov and A.~B.~Zamolodchikov,
  ``On space of integrable quantum field theories,''
Nucl.\ Phys.\ B {\bf 915}, 363 (2017).
[arXiv:1608.05499 [hep-th]].
}

\lref\AsratTZD{
  M.~Asrat, A.~Giveon, N.~Itzhaki and D.~Kutasov,
  ``Holography Beyond AdS,''
Nucl.\ Phys.\ B {\bf 932}, 241 (2018).
[arXiv:1711.02690 [hep-th]].
}

\lref\MaldacenaHW{
  J.~M.~Maldacena and H.~Ooguri,
  ``Strings in AdS(3) and SL(2,R) WZW model 1.: The Spectrum,''
J.\ Math.\ Phys.\  {\bf 42}, 2929 (2001).
[hep-th/0001053].
}

\lref\CavagliaODA{
  A.~Cavaglià, S.~Negro, I.~M.~Szécsényi and R.~Tateo,
  ``$T \bar{T}$-deformed 2D Quantum Field Theories,''
JHEP {\bf 1610}, 112 (2016).
[arXiv:1608.05534 [hep-th]].
}

\lref\SeibergXZ{
  N.~Seiberg and E.~Witten,
  ``The D1 / D5 system and singular CFT,''
JHEP {\bf 9904}, 017 (1999).
[hep-th/9903224].
}
\lref\GiveonFU{
  A.~Giveon, M.~Porrati and E.~Rabinovici,
  ``Target space duality in string theory,''
Phys.\ Rept.\  {\bf 244}, 77 (1994).
[hep-th/9401139].
}

\lref\ApoloQPQ{
  L.~Apolo and W.~Song,
  ``Strings on warped AdS$_{3}$ via $T\bar J $ deformations,''
JHEP {\bf 1810}, 165 (2018).
[arXiv:1806.10127 [hep-th]].
}

\lref\FrolovXZI{
  S.~Frolov,
``$T{\overline T}$, $\widetilde JJ$, $JT$ and $\widetilde JT$ deformations,''
[arXiv:1907.12117 [hep-th]].
}

\Title{} {\centerline{Strings, Symmetric Products,}}
\vskip -1.7in
\Title{} {\centerline{$T \bar{T}$ deformations and Hecke Operators}}

\bigskip
\centerline{\it Akikazu Hashimoto${}^{1}$ and David Kutasov${}^{2}$}
\bigskip
\smallskip
\centerline{${}^{1}$ Department of Physics,  University of Wisconsin, Madison}
\centerline{1150 University Avenue, Madison, WI 53706, USA}
\smallskip
\centerline{${}^2$ EFI and Department of Physics, University of
Chicago} \centerline{5640 S. Ellis Av., Chicago, IL 60637, USA }

\smallskip

\vglue .3cm

\bigskip

\bigskip
\noindent
We derive a formula for the torus partition sum of the symmetric product of $T\bar T$ deformed CFT's, using  previous work on  long strings in (deformed) $AdS_3$, and universality. The result is given by an integral transform of the partition function for the block of the symmetric product, summed over its Hecke transforms, and is manifestly modular invariant. The spectrum is interpretable as a gas of multiply wound long strings with a particular orientation.

\bigskip

\Date{9/19}

In  \HashimotoWCT, we showed\foot{A similar equation was found earlier from a different point of view in \refs{\CardySDV,\DubovskyBMO}.} that the  partition sum of a $T\bar T$ deformed CFT on a torus with modulus $\zeta$ can be expressed as an integral transform of the partition sum of the undeformed CFT (eq. (3.14) in \HashimotoWCT),
\eqn\kernelTT{\ZZ(\zeta, \bar{\zeta},\lambda)={\zeta_2 \over 2 \lambda}
 \int_{\HH_+}{d^2 \tau\over \tau_2^2} e^{-{\pi \over2 \lambda \tau_2} |\tau - \zeta|^2} Z_{\rm cft}(\tau,\bar \tau) \ . }
Here $\HH_+$ is the upper half plane, $Z_{\rm cft}$ is the partition sum of the undeformed CFT, and $\lambda$ is the $T\bar T$ coupling evaluated at the Kaluza-Klein scale $R$, the radius of the spatial circle on which the theory lives (see \HashimotoWCT\ and references therein for further details). 

Our derivation of \kernelTT\ relied on two ideas: the universality of the $T\bar T$ deformation, and the fact that a large class of these theories is obtained by studying the dynamics of long strings on $AdS_3$. In that context, $\tau$ in \kernelTT\ is the modulus of the worldsheet torus. 

We also showed in \HashimotoWCT\ that the string theory point of view is useful for generalizing \kernelTT\ to the case of general combinations of $T \bar T$, $J \bar T$, and $T \bar J$ deformations, and the resulting expressions have the expected modular properties; \eg\ for $T\bar T$ deformed CFT, if the original CFT (with partition sum $Z_{\rm cft}$) is modular invariant, \kernelTT\ satisfies
\eqn\Zmodular{\ZZ\left({a\zeta+b \over c \zeta+d}, {a \bar \zeta+b \over c \bar \zeta+d}, {\lambda\over |c \zeta + d|^2}\right)  = \ZZ(\zeta, \bar \zeta, \lambda),  }
as expected from general considerations  \AharonyBAD. 

In this note we point out that the construction of \HashimotoWCT\ also provides a simple expression for the partition sum of a symmetric product of $N$ $T\bar T$ deformed CFT's, 
\eqn\bbbbb{S^N \MM_\mu=\MM_\mu^N/S_N~.}
Here $\MM_\mu$ is obtained from a CFT $\MM$ by a $T\bar T$ deformation (with coupling $\mu$). Note that \bbbbb\  is different from $\left(S^N \MM \right)_\mu$, the $T\bar T$ deformation of the symmetric product CFT $S^N\MM$. The former, which is often refered to as the single trace $T\bar T$ deformation, corresponds (to first order in the coupling) to adding to the Lagrangian of the symmetric product CFT the term 
\eqn\singletrace{\delta\LL=\mu_s \sum_{i=1}^N T_i \bar T_i~,}
where $T_i$, $\bar T_i$ are the holomorphic and anti-holomorphic components of the stress-tensor of the $i$'th copy of the CFT. The latter, a double trace deformation, corresponds to adding to the Lagrangian the term

\eqn\doubletrace{\delta\LL=\mu_d \sum_{i,j=1}^N T_i\bar T_j=\mu_d T\bar T.}
For the double trace deformation \doubletrace, the analysis of \HashimotoWCT\ applies directly and the partition sum is given by \kernelTT, for the undeformed CFT $S^N\MM$. The single trace deformation \bbbbb\ appears naturally in string theory on $AdS_3$ \refs{\GiveonNIE\GiveonMYJ\AsratTZD-\ChakrabortyKPR}, and we will use this to study it. 

We first review the derivation of \kernelTT\ in \HashimotoWCT, and then extend it to the general case \bbbbb. Consider the bosonic string\foot{To avoid BF violating tachyons we should be studying the superstring, but the difference between the two will, for the most part, not play a role in our discussion. We will comment on the superstring case below.} on 
\eqn\adsthree{AdS_3\times\NN,} 
where $\NN$ is a compact worldsheet CFT, and the $AdS_3$ factor is described by a level $k$ WZW model for the group $SL(2,\IR)$. This background is obtained by starting with an LST background of the form \refs{\AharonyUB,\GiveonZM}
\eqn\rtt{\IR_t\times S^1\times\IR_\phi\times\NN,} 
adding to it $p$ fundamental strings wrapping the $S^1$, and taking the infrared (near-horizon) limit. We will take $p\gg1$, so that the string coupling on $AdS_3$, $g_s^2\sim 1/p$, is small. 

The AdS/CFT correspondence relates string theory on \adsthree\ to a two dimensional CFT. The spectrum of this CFT contains some discrete states corresponding to normalizable string states on \adsthree, and a continuum above a gap, which corresponds to long strings \refs{\MaldacenaUZ\SeibergXZ-\MaldacenaHW}. The theory on one long string is the sigma model on 
\eqn\onestring{\MM=\IR_\phi\times\NN,}
where $\IR_\phi$ describes a scalar field with linear dilaton, whose gradient is fixed by the condition that the total central of \onestring\ is $c_\MM=6k$ \SeibergXZ.\foot{This gradient is in general different from that of the $\IR_\phi$ factor in \rtt.} The theory on $n$ long strings is  $S^n\MM$ (see \eg\ \ChakrabortyMDF\ and references therein). 

As discussed in \refs{\GiveonZM,\GiveonNIE}, one can deform the background \adsthree\ by adding to the worldsheet Lagrangian the term
\eqn\mthree{\delta\LL_{\rm ws}=\alpha J^-\bar J^-,}
where $J^- (\bar J^-)$ is one of the left (right) moving worldsheet $SL(2,\IR)$ currents. In the bulk string theory this modifies the background \adsthree\ to $\MM_3\times\NN$, where $\MM_3$ is the background described \eg\ by eq. (A.4) in \GiveonZM. This background interpolates between \adsthree\ in the infrared region $\phi\to-\infty$, and \rtt\ in the ultraviolet $(\phi\to\infty)$. The former is the near-horizon geometry of $p$ strings in the background \rtt; the latter is the geometry far from the strings. 

The worldsheet deformation \mthree\ corresponds in the spacetime CFT to an irrelevant deformation by an operator of dimension $(2,2)$, that has much in common with $T\bar T$, but is different from it \KutasovXU. In the long string sector, one can show that \mthree\ corresponds to deforming the symmetric product $S^N\MM$ (for $N$ long strings) to \bbbbb\ \refs{\GiveonNIE\GiveonMYJ\AsratTZD-\ChakrabortyKPR,\ChakrabortyMDF}. Thus, calculating the partition sum of long strings on $\MM_3\times\NN$ gives the partition sum of the corresponding symmetric product \bbbbb, with $\MM$ given by \onestring. Universality can then be used to generalize the result to any $\MM$, as in \HashimotoWCT. 

To calculate the partition sum of long strings on $\MM_3\times\NN$, one notes that they correspond to $\delta$-function normalizable states on $\MM_3$, and therefore to compute their spectrum it is enough to study the UV region $\phi\to\infty$, where the background is given by \rtt, and eigenstates of the Hamiltonian are regular plane waves on $\IR_\phi$. The spectrum of energies of these states was studied for the different cases of $T\bar T$ in \refs{\GiveonNIE,\GiveonMYJ}, $J\bar T$ in \refs{\ChakrabortyVJA,\ApoloQPQ}, and the general  combination of deformations in \ChakrabortyMDF. In all these cases, the spectrum was found to be in agreement with a direct field theoretic analysis, for $T\bar T$ that of \refs{\SmirnovLQW,\CavagliaODA}, for $J\bar T$ \ChakrabortyVJA, and for the general case \refs{\LeFlochRUT,\FrolovXZI}.

To study the corresponding torus partition sum, one rotates time to Euclidean signature, and compactifies it together with the spatial coordinate (the $S^1$ in \rtt), on a torus with modulus $\zeta$. Thus, the torus partition sum of long strings is given by the partition sum of string theory on 
\eqn\background{T^2 \times \IR_\phi\times \NN, }
where we only keep states with non-zero winding around the spatial circle. As discussed in \HashimotoWCT, the full partition sum of strings on \background\ is given by 
\eqn\stringZ{\ZZ(\zeta,\bar \zeta, \lambda) = {\zeta_2 \over 2 \lambda} \sum_{m_i, w_i\in Z} \int_{\FF} {d^2 \tau \over \tau_2^2}  e^{-S_{\{m_i,w_i\}}} Z_{\perp}(\tau,\bar{\tau})~,}
where $\FF$ is a fundamental domain of the modular group, $Z_{\perp}$ is the torus partition function of the CFT $\MM$ \onestring, and
\eqn\Smw{S_{\{m_i,w_i\}} = {\pi (w_2 \zeta + w_1)(w_2 \bar \zeta + w_1) \over 2 \lambda \tau_2} \left(\tau - {m_2 \zeta + m_1 \over w_2 \zeta+w_1} \right)
\left(\bar \tau - {m_2 \bar \zeta + m_1 \over w_2 \bar \zeta+w_1} \right) }
is the action\foot{Equation \Smw\ is a re-writing of (3.6) of \HashimotoWCT.} of a linear map from the worldsheet $T^2$ with modulus $\tau$ to the spacetime $T^2$ with modulus $\zeta$. 

The partition sum \stringZ\ includes contributions from all perturbative string states in  the theory. To be more precise, the quantity
\eqn\Xitotal{\ZZ^{(g)}(\zeta, \bar \zeta, \lambda) \equiv \exp\;\ZZ(\zeta, \bar \zeta, \lambda) }
is the trace of $e^{-\beta H+i\sigma p}$, with $\beta\propto\zeta_2$, and $\sigma\propto\zeta_1$,  over all string states, including multi string states constructed out of single string states of arbitrary winding (positive and negative). 

To make the connection with \bbbbb, we would like to restrict the sum to run over multi string states with total string number $N$, constructed out of single string states with positive winding. To do that, we follow  \HashimotoWCT\ and turn on a $B$-field, $\tilde B$,  on the $T^2$ \background. The partition sum \stringZ\ becomes now a function of $\tilde B$ as well as the other variables, and in a suitable normalization of $\tilde B$, is periodic with period one, $\ZZ(\zeta, \bar \zeta, \lambda,\tilde B+1)=\ZZ(\zeta, \bar \zeta, \lambda,\tilde B)$. As discussed in \HashimotoWCT, it can be written as 
\eqn\ZZb{ \ZZ(\zeta, \bar \zeta, \lambda,\eta) = \sum_{N=-\infty}^\infty  \ZZ_N(\zeta,\bar \zeta, \lambda) \eta^N,}
where $\eta=e^{- 2 \pi i \tilde B}$ and 
\eqn\Nmwmw{N = w_1 m_2 - w_2 m_1.}
The integer $N$ \Nmwmw\ is the winding number of the map from the worldsheet torus to the target space one. To restrict the trace to states with a given total string number $N$, we need to pick out the coefficient of $\eta^N$ in $\ZZ^{(g)}=\exp\;\ZZ$ as is defined in  \Xitotal\ and \ZZb. There is an infinite number of such contributions, due to the fact that the winding number $N$ given by   \Nmwmw\ can take both positive and negative values. 

In \HashimotoWCT\ we focused on the contribution $\ZZ_1$ (denoted by $\ZZ$ in \kernelTT) to the partition sum \ZZb. This is the partition sum of single string states with winding one, which via the results of \GiveonNIE\ is a $T\bar T$ deformed CFT of \onestring. In order to generalize the discussion to \bbbbb\ we need to do two things: 
\item{(1)} Look at the coefficient of $\eta^N$ with $N>1$ in the partition sum $\ZZ^{(g)}$.
\item{(2)} Keep only the contributions of states with total winding number $N$, which are combinations of states of positive winding. Examples of contributions that should be included are those of single string states with winding $N$, two string states with windings $N-1$ and $1$, etc. Examples of states that shouldn't contribute are two string states with windings $N+1$ and $-1$, or $N$ and $0$, etc. 

\noindent
To achieve that, we consider the modified partition sum 
\eqn\XiN{\Xi(\zeta, \bar \zeta, \lambda, \eta) \equiv \exp\left[\sum_{N=1}^\infty \eta^N \ZZ_N(\zeta, \bar \zeta, \lambda) \right] }
which differs from $\ZZ^{(g)}$ in that the sum over $N$ runs only over the positive integers, rather than over all integers. 

Our discussion above leads to the conjecture that expanding \XiN\ in a power series in $\eta$, 
\eqn\Xin{\Xi(\zeta, \bar \zeta, \lambda, \eta) = 1 + \sum_{N=1}^\infty \eta^N \Xi_N(\zeta, \bar \zeta, \lambda),}
the coefficient of $\eta^N$, $\Xi_N(\zeta, \bar \zeta, \lambda)$, is the partition sum of the symmetric product \bbbbb\ of $N$ $T\bar T$ deformed CFT's of the form \onestring. 

The argument above is a ``physics proof'' of the formula for the partition sum of \bbbbb, but one can prove directly that it is correct, by using results of previous work \DijkgraafXW. To do that, we examine the partition sum $\ZZ_N$ in \XiN,
\eqn\finalstringZn{\ZZ_N(\zeta,\bar \zeta, \lambda) = {\zeta_2 \over 2 \lambda} \sum_{m_i, w_i|\;N} \int_{\FF} {d^2 \tau \over \tau_2^2}  e^{-S_{\{m_i,w_i\}}} Z_{\perp}(\tau,\bar{\tau}).}
It is convenient to assemble $(w_i,m_i)$ into a matrix 
\eqn\mwmwmatrix{\pmatrix{m_2 & m_1 \cr w_2 & w_1 }.}
The sum in \finalstringZn\ is restricted to matrices \mwmwmatrix\ with determinant $N$ (see \Nmwmw). Each such matrix can be written uniquely as one of the matrices 
\eqn\Tn{ T_N = \left\{\pmatrix{a& b \cr 0 & d }, \qquad a,b,d \in {\bf Z} \qquad ad=N, \qquad 0 \le b < d \right\}}
multiplied from the left by an element of $SL(2,Z)$ \DijkgraafXW. We can use this to replace the sum over $(w_i,m_i)$ in \finalstringZn\ by a restricted sum over matrices of the form \Tn, and trade the sum over elements of $SL(2,Z)$ multiplying these matrices for an extension of the integration region to the upper half plane. This ``unfolding'' is a generalization of the one done in \HashimotoWCT\ for $N=1$. 

Using it and plugging in the values \Tn\ into \finalstringZn, we find that 
\eqn\ZZn{\ZZ_N(\zeta, \bar \zeta, \lambda) =  T_N\left[ \ZZ_1(\zeta,\bar \zeta, \lambda) \right] }
where the Hecke operator $T_N$ acts as 
\eqn\zetagamma{ T_N [\ZZ_1(\zeta,\bar \zeta, \lambda)] = {1 \over N} \sum_{a,b,d} \ZZ_1\left({a \zeta+b \over d}, {a \bar \zeta+b \over d}, {\lambda \over d^2}\right)}
and the sum over $(a,b,d)$ runs over the values in \Tn. Equation \zetagamma\ is the equivalent of (3.5) in \DijkgraafXW.

Plugging \ZZn\ into \XiN\ and using \Xin\ we see that the result we got from our physics arguments agrees with that obtained in \DijkgraafXW\ by studying the contributions to the partition sum\foot{The authors of  \DijkgraafXW\ studied the elliptic genus rather than the partition sum, but the difference is not important for our purposes.} of the various sectors of the symmetric orbifold. 

Some comments are useful at this point:
\item{(1)} The full partition sum \stringZ, and its generalization to non-zero $B$-field, \ZZb, is modular invariant \Zmodular. Since the $B$-field does not transform under modular transformations of the target space torus (see \eg\ \GiveonFU), the Fourier components \ZZb, \finalstringZn\ are modular invariant as well, and the same is true for $\Xi_N$ \Xin. 
\item{(2)} We presented the discussion in the context of string theory on $AdS_3$, where the undeformed CFT $\MM$ that enters the discussion is \onestring, and its partition sum is $Z_\perp$ in \stringZ. As in \HashimotoWCT, we can use the universality of $T\bar T$ deformed CFT to generalize the result to arbitrary unperturbed CFT's, by replacing $Z_\perp$ in \stringZ\ with the partition sum of the undeformed CFT $Z_{\rm cft}$.

Our main conclusion is that the partition function of the symmetric product $S^N {\cal M}_\mu$ is given by $\Xi_N$ defined via \XiN\ and \Xin. It may be useful to make this quantity more explicit for low values of $N$ and to highlight its properites. For $N=2$, we have
\eqn\XiZtwo{
\Xi_2 = {1 \over 2} \ZZ_1^2 + \ZZ_2~.}
We expect \XiZtwo\ to be the partition sum of $S^2\MM_\mu=\left(\MM_\mu\times\MM_\mu\right)/Z_2$.

Using \ZZn, \zetagamma, and the fact that for $N=2$ there are three different $T_N$ \Tn,
\eqn\Ttwo{T_2 = \left\{ \pmatrix{ 2 & 0 \cr 0 & 1}, \
\pmatrix{ 1 & 0 \cr 0 & 2}, \ \pmatrix{ 1 & 1 \cr 0 & 2} \right\} ,}
we can rewrite \XiZtwo\ as 
\eqn\Xii
{\Xi_2(\zeta, \bar \zeta, \lambda)  = {1 \over 2} \ZZ_1(\zeta, \bar \zeta, \lambda) ^2 + {1 \over 2} \ZZ_1(2 \zeta, 2 \bar \zeta, \lambda)+ {1 \over 2} \ZZ_1\left({\zeta \over 2},{\bar \zeta \over 2},{ \lambda \over 4}\right) + {1 \over 2} \ZZ_1\left({\zeta+1 \over 2},{\bar \zeta+1 \over 2}, {\lambda \over 4}\right).}
The first two terms in \Xii\ give the contribution of the untwisted sector, while the last two terms give the contribution of the $Z_2$ twisted sector. In our string theory construction, the latter consists of single string states with winding number two around the spatial circle. Note also that as $\lambda \rightarrow 0$, \Xii\ reproduces the standard expression  for the partition function of a symmetric product CFT (see \eg\ (10) of \KlemmDF).

For $N=3$, we have
\eqn\XiZthree{\eqalign{\Xi_3  &= {1 \over 6} \ZZ_1^3  + \ZZ_1 \ZZ_2 + \ZZ_3 \cr
& =
{1 \over 6} \ZZ_1(\zeta,\bar \zeta, \lambda)^3 \cr
&  
+ {1\over 2}  \ZZ_1\left(2 \zeta, 2\bar \zeta, \lambda\right)\ZZ_1(\zeta,\bar \zeta, \lambda) +
\sum_{i=0}^1
{1 \over 2} \ZZ_1\left({\zeta+ i \over 2}, {\bar\zeta+ i\over 2}, {\lambda \over 4} \right)\ZZ_1(\zeta,\bar \zeta, \lambda) \cr
& + {1\over 3}  \ZZ_1\left(3 \zeta, 3\bar \zeta, \lambda\right)+
    \sum_{i=0}^2  {1 \over 3} \ZZ_1\left({\zeta+ i \over 3}, {\bar\zeta+ i\over 3}, {\lambda \over 9}\right)
}}
where the first terms on each of the last three lines combine into the partition function of the untwised sector, the second term on the third line corresponds to the product of the $Z_2$ twisted and the untwisted sector, and the second term on the last line corresponds to the $Z_3$ twisted sector.

The fact that these terms combine into something interpretable as the Boltzman sum for arbitrary $N$ follows from the arguments in \DijkgraafXW, and in the limit $\lambda\to 0$ the result agrees with \KlemmDF.

For general $N$, from \XiN\ and \Xin\ one sees that $\Xi_N$ contains $\ZZ_N$ with unit coefficient. $\ZZ_N$, in turn, includes a contribution from the Hecke sum \zetagamma\ corresponding to the elements
\eqn\ntwist{\pmatrix{1 & i \cr 0 & N }}
with $0 \le i < N$. These give rise to
\eqn\XiNtwist{\Xi_N^{(\rm twisted)} =  {1 \over N}\sum_{i=0}^{N-1}   \ZZ_1 \left({\zeta+ i \over N}, {\bar\zeta+ i\over N}, {\lambda \over N^2}\right) \ . }
This partition sum can be thought of as describing states living on a circle whose radius is $N$ times larger than the original $R$, but
whose momentum is quantized in units of $1/R$, rather than $1/NR$. 

From the orbifold CFT perspective, these states form the $Z_N$ twisted sector. From the string theory perspective, they are states with winding $N$ around the circle of radius $R$ (and integer momentum). The rest of the contributions to $\Xi_N$ correspond to symmetrized products of various $Z_w$ twisted sectors with total winding number $N$, \ie\ multi string states.

We finish with a few comments:
\item{(1)} Here we discussed the case of single trace $T\bar T$ deformation of a CFT \singletrace, but following \HashimotoWCT\ it is easy to generalize the discussion to the case of general combinations of $T\bar T$, $J\bar T$, $T\bar J$ and $J\bar J$ deformations, for any combination of left and right-moving $U(1)$ currents. 
\item{(2)} We assumed that the canonical partition sum is obtained by compactifying Euclidean time on a circle. This is true when the theory does not have (spacetime) fermions. If it does, the construction of the canonical partition sum is a bit more complicated, since the fermions need to be taken to be anti-periodic around Euclidean time. In string theory it is well known how to implement these boundary conditions (see \eg\ \AtickSI), and we can follow this construction here. We should note that if the theory has spacetime fermions, the partition sum cannot be modular invariant since fermions have half integer $L_0-\bar L_0$, so even before the deformation, a theory that has them is not invariant under the $T$ modular transformation. 
\item{(3)}  Another example (in addition to string theory on $AdS_3$) where the construction of this note is relevant is in the theory on $N$ strings in flat spacetime. As is familiar from matrix string theory \DijkgraafVV, at low energies this theory flows to the CFT $\IR^{8N}/S_N$. If we keep the radius of the circle the $N$ strings wrap finite, we find a theory of the form \bbbbb, with $\MM=\IR^8$. It would be interesting to study it further.

\bigskip\bigskip
\noindent{\bf Acknowledgements:}
We thank S. Dubovsky, A. Giveon, J. Harvey and S. Sethi for discussions. The work of AH is supported
in part by DOE grant DE-SC0017647.  The work of DK is supported in
part by DOE grant DE-SC0009924.  DK thanks the Hebrew University, Tel
Aviv University and the Weizmann Institute for hospitality during part
of this work.

\listrefs
\end